\documentclass{article}
\usepackage{authblk, amsfonts, amsmath, bm, hyperref, multirow}
\usepackage[indent=0pt]{parskip}
\usepackage[a4paper]{geometry}
\usepackage{cleveref}

\usepackage[
    backend = biber,
    style   = ieee,
    ]{biblatex}
\hypersetup{
    colorlinks          = true,
    linkcolor           = blue,
    urlcolor            = magenta,
    bookmarksnumbered   = true,
    citecolor           = red,
    pdftitle            = {On the Continuity Equation in Space-Time Algebra},
    pdfauthor           = {Manuel Beato \& Melvin Arias},
}

\numberwithin{equation}{section}
\allowdisplaybreaks

\title{On the Continuity Equation in Space-Time Algebra: Multivector Waves, Poynting, Diffusion, and a Derivation of Maxwell's Equations by Symmetries}
\author[1]{Manuel Beato V\'asquez}
\author[1, 2]{Melvin Arias Polanco\thanks{melvin.arias@intec.edu.do}}
\affil[1]{Escuela de Física, Facultad de Ciencias, Universidad Autónoma de Santo Domingo, Av. Alma Mater, Santo Domingo 10105, Dominican Republic}
\affil[2]{Laboratorio de Nanotecnología, Área de Ciencias Básicas y Ambientales, Instituto Tecnológico de Santo Domingo, Av. Los Próceres, Santo Domingo 10602, Dominican Republic}
\date{February 2024}

\begin{document}

\maketitle

\begin{abstract}
    \noindent Historically and to date, the continuity equation has served as a consistency criterion for the development of physical theories. Employing Clifford's geometric algebras, a system of continuity equations for a generalised multivector of the space-time algebra (STA) is constructed. Associated with this continuity system, a system of wave equations is constructed, the Poynting multivector is defined, and decoupling conditions are determined. The diffusion equation is explored from the continuity system, where it is found that for decoupled systems with constant or explicitly-dependent diffusion coefficients the absence of external vector sources implies a loss in the diffusion equation structure being transformed to Helmholtz-like or wave systems. From the symmetry transformations that make the continuity equations system's structure invariant, a system with the structure of Maxwell's field equations is derived. The Maxwellian system allows for the construction of potentials and fields directly linked with the continuity of a generalised multivector in STA. The results found are consistent with the classical electromagnetic theory and hydrodynamics.
\end{abstract}

\section{Introduction}

The continuity equation (C.E.) is a first-order partial differential equation that describes the transport in time of a physical quantity through a region of space in terms of its density, flux, and an external source (or sink) which increases (or decreases) the quantity from the system continuously. If the source is null, then the transported physical quantity is said to be conservative, and the C.E. represents the principle of local conservation of said quantity \cite{KipThorne}. In this form, the C.E. permeates virtually all branches of physics (\cref{Branch and physical quantity}).

An iconic case where the continuity equation played a fundamental role in the development of a physical theory occurred when J. C. Maxwell noticed the inconsistency between the primitive field equations of the electromagnetic theory and the C.E. for the electric charge (a principle of conservation well established at the time). Maxwell is capable of solving said inconsistency by introducing the now renowned displacement `current'. Thanks to this addition, he was able to derive wave equations for the electric and magnetic fields, subsequently discovering that their speed of propagation is precisely the speed of light \cite{Jackson}. Another instance was in the development of the quantum-relativistic equation by P. A. M. Dirac following previous inconsistent attempts to reconcile both theories under a unified model. One such attempt is the Klein-Fock-Gordon equation, which upon its reduction to a C.E. yields negative energy eigenvalues and a non-positive definite probability density. Dirac, using a matrix analysis and ensuring consistency with the C.E., constructed what we today know as Dirac's gamma matrices, algebra, and equation. Thus, allowing for a description of spin-$1/2$ massive particles in the presence of external fields while taking into account relativistic effects \cite{Sakurai2020}. 

\begin{table}[h]
    \centering
    \footnotesize
    \begin{tabular}{|c|l|}
    \hline
    \multirow{4}{*}{\centering Continuum Mechanics}     & Mass \\
                                                        & Energy \\
                                                        & Linear momentum \\
                                                        & Angular momentum \\
    \hline 
    \multirow{3}{*}{\centering Electrodynamics}         & Electric charge \\
                                                        & Electromagnetic energy \\
                                                        & Electromagnetic momentum \\
    \hline
    \multirow{4}{*}{\centering Thermal Physics}         & Particle number \\
                                                        & Thermal energy \\
                                                        & Entropy \\
                                                        & Probability (Liouville, Fokker-Planck) \\
    \hline
    \multirow{2}{*}{\centering Quantum Mechanics}       & Probability (Schrödinger, Dirac) \\
                                                        & Spin angular momentum \\
    \hline   
    \end{tabular}
    \caption{Conservative physical quantities.}
    \label{Branch and physical quantity}
\end{table}

Undoubtedly, the C.E. has served as a consistency criterion for the development of physical theories. A natural language to express and study the C.E. are W. K. Clifford's geometric algebras (GA). In physics, GA have enjoyed a vast popularisation since their `rediscovery' by W. E. Pauli and Dirac in the 1920's, and D. Hestenes in the 1960's. By design, GA establishes itself as a holistic mathematical framework capable of algebraically manipulating multi-dimensional objects, operating on them with a clear geometric interpretation, and nesting within itself several sub-algebras such as complex numbers, quaternions, four-vectors, and spinors \cite{JLasenby2000, Gull1993, Lasenby2017, Hestenes2002, Baylis1992}. As a result, GA finds numerous applications not only in physics, but also in mathematics, computer science, and engineering \cite{Hitzer2024, Breuils2022}.

Axiomatically, a geometric algebra $\mathcal{G}(p, q)$ is a graded and associative algebra of dimension $2^{p+q}$ defined by a vector space $\mathcal{V}^{p+q}$ over the field of real numbers $\mathbb{R}$ \cite{Macdonald2002, Arthan2006, Macdonald2017, HITZER2012}. Elements of a GA are called multivectors and operate through the so-called geometric product. The geometric product---denoted by juxtaposition---is closed, distributive, `well behaved' with scalars, invertible, and associative. For vectors, the geometric product is linked to the inner product of the vector space and H. G. Grassmann's outer product via the so-called fundamental identity 

\begin{equation}
    \gamma_{\mu} \gamma_{\nu} = \gamma_{\mu} \cdot \gamma_{\nu} + \gamma_{\mu} \wedge \gamma_{\nu}
\end{equation}

In general, for two homogeneous multivectors $\Psi_{\alpha} = \langle \Psi_{\alpha} \rangle_{\alpha}$ and $\Phi_{\beta} = \langle \Phi_{\beta} \rangle_{\beta}$ of grades $\alpha$ and $\beta$ respectively, the inner and outer products are defined in terms of the geometric product as \cite{Hestenes1984, Doran-Lasenby2003, Doran1994}:

\begin{align}
    \Psi_{\alpha} \cdot \Phi_{\beta} &= \langle \Psi_{\alpha} \Phi_{\beta} \rangle_{| \alpha - \beta |} \\
    \Psi_{\alpha} \wedge \Phi_{\beta} &= \langle \Psi_{\alpha} \Phi_{\beta} \rangle_{\alpha + \beta}
\end{align}

Consider the geometric algebra $\mathcal{G}(1,3)$, known as the space-time algebra (STA), generated by the orthonormal basis vectors $\left\{ \gamma_0, \gamma_1, \gamma_2, \gamma_3 \right\}$ which satisfy \cite{Hestenes2015, Hestenes2003, Baylis2004}:

\begin{equation}
    \label{STA inner product}
    \gamma_{\mu} \cdot \gamma_{\nu} = \frac{1}{2} \left( \gamma_{\mu}\gamma_{\nu} + \gamma_{\nu}\gamma_{\mu} \right) = \eta_{\mu\nu}
\end{equation}

where $\eta_{\mu\nu} = \text{diag}(+1, -1, -1, -1)$ is Minkowski's metric tensor. The relations with the dual basis are

\begin{equation}
    \gamma_\mu = \eta_{\mu\nu} \gamma^\nu, \qquad 
    \gamma_{\mu} \cdot \gamma^{\nu} = \delta_{\mu}^{\nu}
\end{equation}

where $\delta^\nu_\mu = \eta^{\nu \alpha} \eta_{\alpha \mu}$ is the symmetric Kronecker delta tensor. A key corollary of \cref{STA inner product} is that orthogonal vectors anticommute under the geometric product: $\gamma_\mu \gamma_\nu = - \gamma_\nu \gamma_\mu$, $\forall \mu \neq \nu$. Henceforth we adopt Einstein's summation convention; Greek indices run from 0 to 3, and Latin indices from 1 to 3 (except when explicitly stated otherwise); and we make use of the so-called `natural units' where $c=1$. Let $I = \gamma_0 \gamma_1 \gamma_2 \gamma_3$ be the unit pseudo-scalar of $\mathcal{G}(1,3)$ with properties

\begin{equation}
    II = -1, \qquad 
    I \gamma_{\mu} = - \gamma_{\mu} I
\end{equation} 

Let us introduce the space-time split given by the sub-algebra of physical space (APS), $\mathcal{G}^+(1, 3) \simeq \mathcal{G}(3,0)$, generated by the bivectors

\begin{equation}
    \bm{\sigma}_i = \gamma_i \gamma_0, \qquad 
    \bm{\sigma}_i \bm{\sigma}_j = \delta_{ij} + \varepsilon_{ijk}\,I\bm{\sigma}_k
\end{equation}

where $\varepsilon_{ijk}$ is the antisymmetric Levi-Civita tensor and such that the relative vectors are defined\footnote{Another property we shall use numerously but implicitly is the commutation of the pseudoscalar with the relative vectors: $I \bm{\sigma}_i = \bm{\sigma}_i I$.}:

\begin{equation}
    \bm{\psi}_{\alpha} =  \psi^i_{\alpha} \, \bm{\sigma}_i, \qquad \bm{\nabla} = \bm{\sigma}_i \, \partial_i
\end{equation}

In STA, the vector derivative operator with respect to space-time position $x = x^\mu \gamma_\mu = (t + \mathbf{x}) \gamma_0$ is defined as

\begin{equation}
\label{nabla}
    \nabla = \gamma^{\mu}\,\partial_{\mu} = \left( \partial_t - \bm{\nabla} \right)\gamma_0
\end{equation}

For brevity, we will refer to \cref{nabla} as the nabla operator. Let us also define the structure of a generalised multivector of STA as:

\begin{equation}
\label{Psi}
\begin{split}
    \Psi &= \langle \Psi \rangle_0 + \langle \Psi \rangle_1 + \langle \Psi \rangle_2 + \langle \Psi \rangle_3 + \langle \Psi \rangle_4 \\
    &= \psi_{2}^0 + \psi_{1}^{\mu} \, \gamma_\mu + \psi_{2}^{i} \, \gamma_i\gamma_0 + \psi_{3}^{i} \, I \gamma_i \gamma_0 + \psi_{4}^{\mu} \, I \gamma_{\mu} + \psi_{3}^0 \, I \\
    &= \psi^0_2 + \left( \psi_{1}^{0} + \bm{\psi}_{1} \right)\gamma_0 + \bm{\psi}_{2} + \bm{\psi}_{3} \, I + \left( \psi_{4}^{0} + \bm{\psi}_{4} \right) I \gamma_0 + \psi^0_{3} \, I
\end{split}
\end{equation}

where\footnote{Throughout the document: $\alpha, \beta = 1, 2, 3, 4$.} $\psi_{\alpha}^{\mu} = \psi_{\alpha}^{\mu} \left( x^\mu \right)$ are scalar functions of the coordinates of $x$. Respectively, they constitute the scalar $\langle \Psi \rangle_0 = \psi^0_2$, vector $\langle \Psi \rangle_1 = \left( \psi_{1}^{0} + \bm{\psi}_{1} \right)\gamma_0$, timelike bivector $\langle \Psi \rangle_{2t} = \bm{\psi}_2$, spacelike bivector\footnote{The integral bivector part of $\Psi$ is $\langle \Psi \rangle_{2} = \langle \Psi \rangle_{2t} + \langle \Psi \rangle_{2s} = \bm{\psi}_{2} + \bm{\psi}_{3}I$.} $\langle \Psi \rangle_{2s} = \bm{\psi}_3 I$, pseudo-vector $\langle \Psi \rangle_3 = \left( \psi_{4}^{0} + \bm{\psi}_{4} \right) I \gamma_0$, and pseudo-scalar part $\langle \Psi \rangle_4 = \psi^0_3 I$ of the multivector $\Psi$. For the generalised multivector defined in \cref{Psi} the following conjugation operations are defined:

\begin{itemize}
    \item[$\bullet$] Inverse conjugation or reversal $\left(\gamma_\alpha \wedge \gamma_\beta \wedge \cdots \wedge \gamma_\mu \wedge \gamma_\nu\right)^{\sim} = \gamma_\nu \wedge \gamma_\mu \wedge \cdots \wedge \gamma_\beta \wedge \gamma_\alpha$,

    \begin{equation}
    \widetilde{\Psi} = \left(\psi_{1}^{0} + \bm{\psi}_{1}\right)\gamma_0 + \left(\psi^0_2 - \bm{\psi}_{2}\right) + \left(\psi^0_{3} - \bm{\psi}_{3}\right)I - \left(\psi_{4}^{0} + \bm{\psi}_{4}\right)I\gamma_0
    \end{equation}

    \item[$\bullet$] Space conjugation $\Psi^* = \gamma_0 \Psi \gamma_0$,

    \begin{equation}
    \Psi^* = \left(\psi_{1}^{0} - \bm{\psi}_{1}\right)\gamma_0 + \left(\psi^0_2 - \bm{\psi}_{2}\right) - \left(\psi^0_{3} - \bm{\psi}_{3}\right) I - \left(\psi_{4}^{0} - \bm{\psi}_{4}\right) I\gamma_0
    \end{equation}
    
    \item[$\bullet$] Hermitian conjugation $\Psi^\dagger = \widetilde{\Psi}^*$, 

    \begin{equation}
    \label{psi dagger}
    \Psi^\dagger = \left(\psi_{1}^{0} - \bm{\psi}_{1}\right)\gamma_0 + \left(\psi^0_2 + \bm{\psi}_{2}\right) - \left(\psi^0_{3} + \bm{\psi}_{3}\right) I + \left(\psi_{4}^{0} - \bm{\psi}_{4}\right) I\gamma_0
    \end{equation}
\end{itemize}

It is assumed the invariance of the scalar functions $\psi^\mu_\alpha$ under all conjugations, that they satisfy Clairaut-Schwarz's theorem, and commute with the basis vectors and the pseudoscalar. Attending to these definitions, the reversal of the nabla \cref{nabla} indicates that the partial derivatives act to the left, $\widetilde{\nabla} = \overleftarrow{\nabla}$, while hermitian conjugation yields

\begin{equation}
\label{nabla dagger}
    \nabla^\dagger = \left(\overleftarrow{\partial_t} + \overleftarrow{\bm{\nabla}}\right)\gamma_0
\end{equation}

With this preamble, in the present paper, we propose to employ the mathematical framework of Clifford's geometric algebras to study systems of continuity equations for generalised multivectors in STA. Our starting point in \cref{Sec Continuity} will consist of establishing an equation of structure $\nabla \Psi = \Phi$ and identifying the systems of continuity equations that emerge from each component of the multivector defined by \cref{Psi}. Then, we will construct a system of wave equations and define the generalised Poynting multivector (along with its continuity) associated with the continuity system of the generalised multivector. STA contains APS which further contains the quaternion algebra, $G^+(3, 0) \simeq \mathbb{H}$. So, our approach with generalised multivectors and notably the formalism of STA distinguishes this work from similar ones using a quaternionic formalism \cite{Arbab2010, ARBAB2021}, and with explorations of the Poynting vector restricted to an electromagnetic treatment in GA \cite{Sugon2008, Castilla2009} and hydrodinamical without GA \cite{Scofield2014}. In \cref{Sec Desacoplamientos Lineales} we will explore some decoupling cases for the different parts of $\Psi$ which are coupled with each other through the continuity system. Subsequently, in \cref{Sec Symmetries} we shall determine transformation symmetries for the functions involved in the continuity system such that the system of continuity equations remains invariant in structure. From these symmetries we plan to derive a system of equations with the same structure as Maxwell's field equations. Also, together with it, see an application of the wave equations system, the Poynting multivector, and the decoupling conditions for the fields and potentials to be determined. Alternative derivations of Maxwell's equations from the C.E. can be found in \cite{Burns2019, Macdonald2019} by means of retarded potentials and the Poincaré lemma in GA; also in \cite{Heras2007, Heras2009} using Jefimenko's equations and retarded tensor fields without GA. Finally, in \cref{Sec Diffusion} we will employ the continuity system to study the diffusion equation and the equations coupled to it.

\section{Continuity Equations} \label{Sec Continuity}

Consider the equation

\begin{equation}
\label{nabla psi = phi}
     \nabla \Psi = \Phi
\end{equation}

where $\Phi = \Phi(x^\mu)$ is another multivector with the structure of \cref{Psi}. The geometric product between the nabla and the multivector $\Psi$ yields

\begin{equation}
\begin{split}
\label{nabla Psi}
    \nabla \Psi =& \; \left( \partial_t \psi_1^0 + \bm{\nabla} \cdot \bm{\psi}_1 \right) \\
                &+ \left( \partial_t \psi_2^0 + \bm{\nabla} \cdot \bm{\psi}_2 - \partial_t\bm{\psi}_2 - \bm{\nabla}\psi_2^0 - I \, \bm{\nabla} \wedge \bm{\psi}_3 \right) \gamma_0 \\
                &- \left( \partial_t\bm{\psi}_1 + \bm{\nabla}\psi_1^0 + I \, \bm{\nabla} \wedge \bm{\psi}_4 \right) \\
                &+ \left( \partial_t\bm{\psi}_4 + \bm{\nabla}\psi_4^0 - I \, \bm{\nabla} \wedge \bm{\psi}_1 \right) I \\
                &+ \left( - \partial_t \psi^0_3 - \bm{\nabla} \cdot \bm{\psi}_3 + \partial_t\bm{\psi}_3 + \bm{\nabla}\psi_3^0 - I \, \bm{\nabla} \wedge \bm{\psi}_2 \right) I \gamma_0 \\
                &- \left( \partial_t \psi_4^0 + \bm{\nabla} \cdot \bm{\psi}_4 \right) I
\end{split}
\end{equation}

The terms in \cref{nabla Psi} have been organised by parenthesis into scalar, vector, timelike bivector, spacelike bivector, pseudovector, and pseudoscalar part, respectively. By hypothesis, $\Phi$ has structure \cref{Psi}, so matrix notation allows us to identify the following pair of systems of equations which follow from the components of \cref{nabla psi = phi}. On one hand, the scalar structure:

\begin{equation}
\label{continuity psi}
    \partial_t \left(
    \begin{array}{c}
    \psi_1^0 \\
    \psi_2^0 \\
    \psi_3^0 \\
    \psi_4^0 \\
    \end{array} \right)
    + \bm{\nabla} \cdot \left(
    \begin{array}{c}
    \bm{\psi}_1 \\
    \bm{\psi}_2 \\
    \bm{\psi}_3 \\
    \bm{\psi}_4 \\
    \end{array} \right)
    = \left(
    \begin{array}{c}
    \phi_2^0 \\
    \phi_1^0 \\
    - \phi_4^0 \\
    - \phi_3^0
    \end{array} \right)
\end{equation}

And on the other hand the vector structure:

\begin{equation}
\label{field psi}
    \partial_t \left(
    \begin{array}{c}
    \bm{\psi}_1 \\
    \bm{\psi}_2 \\
    \bm{\psi}_3 \\
    \bm{\psi}_4 \\
    \end{array} \right)
    + \bm{\nabla} \left(
    \begin{array}{c}
    \psi_{1}^{0} \\
    \psi_{2}^0 \\
    \psi_{3}^0 \\
    \psi_{4}^{0} \\
    \end{array} \right)
    - I \, \bm{\nabla} \wedge \left(
    \begin{array}{c}
    - \bm{\psi}_4 \\
    - \bm{\psi}_3 \\
    \bm{\psi}_2 \\
    \bm{\psi}_1 \\
    \end{array} \right)
    = \left(
    \begin{array}{c}
    - \bm{\phi}_2 \\
    - \bm{\phi}_1 \\
    \bm{\phi}_4 \\
    \bm{\phi}_3
    \end{array} \right)
\end{equation}

The system of \cref{continuity psi} clearly possesses the canonical structure of the (scalar) continuity equation with existence of non-zero sources given by $\Phi$. To observe that the system of \cref{field psi} has a continuity structure as well, but one of vector character, we use the dual identity

\begin{equation}
\label{rot and div duality}
    I \, \bm{\nabla} \wedge \bm{\psi}_{\alpha} = \bm{\nabla} \cdot \left( I \bm{\psi}_{\alpha} \right)
\end{equation}

and define the second-rank tensors $\left( \bm{\mathsf{T}}_{\alpha} \right)^{ij}$ such that its components are connected with the $\psi_{\alpha}^0$ through the restriction

\begin{equation}
\label{Tensor T and psi relation}
    \partial_i T_{\alpha}^{ij} = \partial_j\psi_{\alpha}^0, \qquad \bm{\nabla} \cdot \bm{\mathsf{T}}_{\alpha} = \bm{\nabla} \psi_{\alpha}^0
\end{equation}

With these results, \cref{continuity psi,field psi} compact into a visibly single continuity structure: 

\begin{equation}
\label{tensor continuity psi}
    \partial_t \left(
    \begin{array}{c}
    \psi_1^0 \\
    \psi_2^0 \\
    \psi_3^0 \\
    \psi_4^0 \\
    \bm{\psi}_1 \\
    \bm{\psi}_2 \\
    \bm{\psi}_3 \\
    \bm{\psi}_4
    \end{array} \right)
    + \bm{\nabla} \cdot \left(
    \begin{array}{c}
    \bm{\psi}_1 \\
    \bm{\psi}_2 \\
    \bm{\psi}_3 \\
    \bm{\psi}_4 \\
    \bm{\mathsf{T}}_1 + I \, \bm{\psi}_4 \\
    \bm{\mathsf{T}}_2 + I \, \bm{\psi}_3 \\
    \bm{\mathsf{T}}_3 - I \, \bm{\psi}_2 \\
    \bm{\mathsf{T}}_4 - I \, \bm{\psi}_1
    \end{array} \right)
    = \left(
    \begin{array}{c}
    \phi_2^0 \\
    \phi_1^0 \\
    - \phi_4^0 \\
    - \phi_3^0 \\
    - \bm{\phi}_2 \\
    - \bm{\phi}_1 \\
    \bm{\phi}_4 \\
    \bm{\phi}_3
    \end{array} \right)
\end{equation}

\subsection{Wave Equations}

By applying from the left the nabla operator to \cref{nabla psi = phi} we obtain the second-order differential equation

\begin{equation}
\label{wave Psi = nabla Phi}
    \square \Psi = \nabla\Phi
\end{equation}

where $\square = \nabla\nabla = \nabla\cdot\nabla = \eta^{\mu\nu}\partial^2_{\mu\nu}$ is the d'Alembertian operator. The multivector $\nabla \Phi$ will have by structure \cref{nabla Psi}, so by separating the components of \cref{wave Psi = nabla Phi} we will get two systems of equations with the form of \cref{continuity psi,field psi}. Defining the tensors $(\bm{\mathsf{N}}_{\alpha})^{ij}$ restricted to the analogous condition of \cref{Tensor T and psi relation} but connected to the $\phi^0_\alpha$, the following system of wave equations is then obtained:

\begin{equation}
\label{tensor continuity phi}
    \partial_t \left(
    \begin{array}{c}
    \phi_1^0 \\
    \phi_2^0 \\
    \phi_3^0 \\
    \phi_4^0 \\
    \bm{\phi}_1 \\
    \bm{\phi}_2 \\
    \bm{\phi}_3 \\
    \bm{\phi}_4
    \end{array} \right)
    + \bm{\nabla} \cdot \left(
    \begin{array}{c}
    \bm{\phi}_1 \\
    \bm{\phi}_2 \\
    \bm{\phi}_3 \\
    \bm{\phi}_4 \\
    \bm{\mathsf{N}}_1 + I \, \bm{\phi}_4 \\
    \bm{\mathsf{N}}_2 + I \, \bm{\phi}_3 \\
    \bm{\mathsf{N}}_3 - I \, \bm{\phi}_2 \\
    \bm{\mathsf{N}}_4 - I \, \bm{\phi}_1
    \end{array} \right)
    = \square \left(
    \begin{array}{c}
    \psi_2^0 \\
    \psi_{1}^{0} \\
    - \psi_{4}^{0} \\
    - \psi_3^0 \\
    - \bm{\psi}_2 \\
    - \bm{\psi}_1 \\
    \bm{\psi}_4 \\
    \bm{\psi}_3
    \end{array} \right)
\end{equation}

The scalar continuity system of \cref{continuity psi} shows the direct coupling between the vector components $(\psi_{1}^{0}, \bm{\psi}_1)$, the pseudovector components $(\psi_{4}^{0}, \bm{\psi}_4)$, between the scalar and timelike bivector $(\psi_2^0, \bm{\psi}_2)$, and between the pseudoscalar and spacelike bivector $(\psi_3^0, \bm{\psi}_3)$. We shall call these relations the `first couplings'. The first couplings remain in the vector continuity system of \cref{field psi}, and additionally the relative pseudovector with the vector $(\psi_{1}^{0}, \bm{\psi}_1, \bm{\psi}_4)$, the relative vector with the pseudovector $(\psi_{4}^{0}, \bm{\psi}_4, \bm{\psi}_1)$, and the timelike and spacelike bivectors with each scalar $(\psi_2^0, \bm{\psi}_2, \bm{\psi}_3)$, $(\psi_3^0, \bm{\psi}_3, \bm{\psi}_2)$ are coupled. For consistency, we call these the `second couplings'. An analogous argument applies to the components of the source $\Phi$. The cost for decoupling---first or second--- is a transition to a higher-order system of equations, as can be seen in \cref{tensor continuity phi} where each component $\psi_{\alpha}^{\mu}$ satisfies a non-homogeneous wave equation independently of any other component of $\Psi$. In \cref{Sec Desacoplamientos Lineales} we will explore some decoupling cases.

\subsection{Poynting Multivector}

Attending to the conjugations \cref{psi dagger,nabla dagger}, the hermitian conjugation of \cref{nabla psi = phi} is $(\nabla \Psi)^\dagger = \Psi^\dagger \nabla^\dagger = \Phi^\dagger$. The well-known method to derive the continuity equation from Schrödinger, Pauli, Klein-Fock-Gordon, and Dirac's equation

\begin{equation}
\begin{split}
\label{Dirac's current continuity}
    (\nabla \Psi)^\dagger(\gamma_0 \Psi) + (\gamma_0 \Psi)^\dagger(\nabla \Psi) &= \Phi^\dagger(\gamma_0 \Psi) + (\gamma_0 \Psi)^\dagger \Phi \\
    \partial_\mu \left( \Psi^\dagger \gamma_0 \gamma^\mu \Psi \right) &= \Phi^\dagger \gamma_0 \Psi + \Psi^\dagger \gamma_0 \Phi \\
    \nabla \cdot J &= \gamma_0 \frac{1}{2} \left( \widetilde{\Phi} \Psi + \widetilde{\Psi} \Phi \right)
\end{split}
\end{equation}

allows us to define

\begin{equation}
    \label{J tensor vectors}
    J^\mu = \frac{1}{2} \left( \Psi^\dagger \gamma_0 \gamma^\mu \Psi \right)
\end{equation}

\Cref{J tensor vectors} defines four multivectors which may be associated with a second-rank tensor, $E^{\mu \nu} = J^\mu \cdot \gamma^\nu$. Let us define the Poynting multivector as

\begin{equation}
\label{Poynting multivector}
    S = \gamma_0 J^0 = \frac{1}{2} \gamma_0 \Psi^\dagger \Psi
\end{equation}

In terms of the components of $\Psi$, it has the explicit form

\begin{equation}
\begin{split}
\label{Poynting multivector components}
    S = \; & \left( \psi^0_{1}\psi^0_{2} + \bm{\psi}_1 \cdot \bm{\psi}_2 + \psi^0_{3}\psi^0_{4} + \bm{\psi}_3 \cdot \bm{\psi}_4 \right) \\
        &+ \frac{1}{2} \left[ (\psi^0_{1})^2 + \bm{\psi}_1 \cdot \bm{\psi}_1 + (\psi^0_{2})^2 + \bm{\psi}_2 \cdot \bm{\psi}_2 + (\psi^0_{3})^2 + \bm{\psi}_3 \cdot \bm{\psi}_3 + (\psi^0_{4})^2 + \bm{\psi}_4 \cdot \bm{\psi}_4 \right] \gamma_0 \\
        &+ \left( \psi^0_{1} \bm{\psi}_1 - \psi^0_{2} \bm{\psi}_2 - \psi^0_{3} \bm{\psi}_3 + \psi^0_{4} \bm{\psi}_4 + I \, \bm{\psi}_1 \wedge \bm{\psi}_4 - I \, \bm{\psi}_2 \wedge \bm{\psi}_3 \right) \gamma_0 \\
        &+ \left( \psi^0_{1}\psi^0_{3} + \bm{\psi}_1 \cdot \bm{\psi}_3 - \psi^0_{2}\psi^0_{4} - \bm{\psi}_2 \cdot \bm{\psi}_4 \right) I
\end{split}
\end{equation}

The terms in \cref{Poynting multivector components} have been organised by parenthesis with respect to the `canonical' structure of the multivector \cref{Psi}, $S = S^0_2 + (S^0_1 + \bm{S}_1)\gamma_0 + S^0_3 \, I$. Interestingly, the Poynting multivector lacks bivector and pseudovector parts, $\langle S \rangle_2 = \langle S \rangle_3 = 0$. Let us establish the equation,

\begin{equation}
\label{Poynting continuity}
    \nabla \cdot S = W
\end{equation}

Then, according to the general structure of \cref{tensor continuity psi}, the continuity system which follows from \cref{Poynting continuity} is:

\begin{equation} 
\begin{split}
\label{Poynting continuity components}
    \partial_t S^0_1 + \bm{\nabla} \cdot \bm{S}_1 &= \langle W \rangle_0 \\
    -\nabla S^0_3 &= \langle W \rangle_3
\end{split}
\end{equation}

No second couplings are present; only the first coupling of $S_1^\mu$.

\subsubsection{Bivector \& Vector Cases} \label{Poynting Cases}

As a special case, consider a multivector for which $\psi_1^\mu = \psi_4^\mu = 0$ such that 

\begin{equation}
\label{Psi pure bivector}
    \Psi = (\psi^0_2 + \bm{\psi}_{2}) + (\psi^0_{3} + \bm{\psi}_{3}) I
\end{equation}

The associated Poynting multivector has a purely vector structure

\begin{equation}
    S = \frac{1}{2} \left[ (\psi^0_{2})^2 + \bm{\psi}_2 \cdot \bm{\psi}_2 + (\psi^0_{3})^2 + \bm{\psi}_3 \cdot \bm{\psi}_3 \right] \gamma_0 - \left( \psi^0_{2} \bm{\psi}_2 + \psi^0_{3} \bm{\psi}_3 + I \, \bm{\psi}_2 \wedge \bm{\psi}_3 \right) \gamma_0
\end{equation}

I.e., $S = \left(S^0_1 + \bm{S}_1\right)\gamma_0$. Correspondingly, the continuity system \cref{Poynting continuity components} reduces to the single equation $\partial_\mu S_1^\mu = \langle W \rangle_0$. Now, consider a multivector for which $\psi_2^\mu = \psi_3^\mu = 0$,

\begin{equation}
\label{Psi pure vector}
    \Psi = \left(\psi_{1}^{0} + \bm{\psi}_{1}\right)\gamma_0 + \left(\psi_{4}^{0} + \bm{\psi}_{4}\right) I\gamma_0
\end{equation}

Poynting multivector:

\begin{equation}
\label{S for vector Psi}
    S = \frac{1}{2} \left[ (\psi^0_{1})^2 + \bm{\psi}_1 \cdot \bm{\psi}_1 + (\psi^0_{4})^2 + \bm{\psi}_4 \cdot \bm{\psi}_4 \right] \gamma_0 + \left( \psi^0_{1} \bm{\psi}_1  + \psi^0_{4} \bm{\psi}_4 + I \, \bm{\psi}_1 \wedge \bm{\psi}_4 \right) \gamma_0
\end{equation}

That is, $S\left(\psi_2^\mu = \psi_3^\mu = 0\right)$ has a purely vector structure, just like $S\left(\psi_1^\mu = \psi_4^\mu = 0\right)$. Thus, their reduced continuity system will share the same structure.

\section{`Linear' Decoupling} \label{Sec Desacoplamientos Lineales}

Because of the independence and symmetry between the continuity equations regarding the second couplings, $\left(\psi_1^{\mu}, \psi_4^{\mu}\right)$ \& $\left(\psi_2^{\mu}, \psi_3^{\mu}\right)$, consider the simplified system of \cref{tensor continuity psi}:

\begin{align}
    \label{coupled system scalar alpha}
    \partial_t \psi_{\alpha}^0 + \bm{\nabla} \cdot \bm{\psi}_{\alpha} &= \phi_{\alpha} \\
    \label{coupled system scalar beta}
    \partial_t \psi_{\beta}^0 + \bm{\nabla} \cdot \bm{\psi}_{\beta} &= - \phi_{\beta} \\
    \label{coupled system vector alpha}
    \partial_t \bm{\psi}_{\alpha} + \bm{\nabla} \psi^0_{\alpha} + I \, \bm{\nabla} \wedge \bm{\psi}_{\beta} &= - \bm{\phi}_{\alpha} \\
    \label{coupled system vector beta}
    \partial_t \bm{\psi}_{\beta} + \bm{\nabla} \psi^0_{\beta} - I \, \bm{\nabla} \wedge \bm{\psi}_{\alpha} &= \bm{\phi}_{\beta}
\end{align}

Where it is understood that if $\alpha = 1$ then $\beta = 4$, and correspondingly if $\alpha = 2$ then $\beta = 3$. The source terms $\phi_{\alpha}, \bm{\phi}_{\alpha}$, etc. are given by \cref{tensor continuity psi} although for now we are not interested in their form. Let us proceed to determine some cases for `linear' decoupling---for which we simply mean that we wish to decouple any one of the functions $\psi^\mu_\alpha$ and $\psi^\mu_\beta$ from one another in the system \crefrange{coupled system scalar alpha}{coupled system vector beta} without necessarily recurring to the wave system \cref{tensor continuity phi} or the trivial case. 

\subsection{Mutual Decoupling}

The simplest case of linear decoupling consists of imposing on both relative vectors the mutual restriction $\bm{\nabla} \wedge \bm{\psi}_{\alpha, \beta} = 0$. Whence,

\begin{equation}
    \label{desacoplamiento simultaneo}
    \bm{\psi}_{\alpha, \beta} = \bm{\nabla} \Gamma_{\alpha, \beta}
\end{equation}

where the $\Gamma_{\alpha, \beta} = \Gamma_{\alpha, \beta}(x^\mu)$ are scalar functions, and the index ($\alpha, \beta$) denotes that the above equations apply for both functions simultaneously. With the mutual simultaneous decoupling of \cref{desacoplamiento simultaneo}, the system of \crefrange{coupled system scalar alpha}{coupled system vector beta} reduces to:

\begin{equation}
\begin{split}
    \partial_t \psi_{\alpha, \beta}^0 + \bm{\nabla}^2 \Gamma_{\alpha, \beta} &= \pm \phi_{\alpha, \beta} \\
    \bm{\nabla} \left( \psi^0_{\alpha, \beta} + \partial_t \Gamma_{\alpha, \beta} \right) &= \mp \bm{\phi}_{\alpha, \beta}
\end{split}
\end{equation}

with $\bm{\nabla}^2 = \bm{\nabla} \cdot \bm{\nabla} = \delta_{ij} \partial_{ij}^2$ the Laplacian operator of APS. Thus, imposing the restriction of \cref{desacoplamiento simultaneo} simultaneously to both relative vectors $\bm{\psi}_{\alpha}$ and $\bm{\psi}_{\beta}$ decouples them in two pairs of equations with the same structure but independent of each other.

\subsection{Unilateral Decoupling} \label{Subsection unilateral decoupling}

Suppose we now impose a restriction to just one of the terms, say $\psi^{\mu}_{\beta}$ with respect to $\psi^{\mu}_{\alpha}$. Decoupling is achieved with the condition,

\begin{equation}
\label{psi_4 desacoplamiento}
    \psi_{\beta}^\mu = \partial^\mu \Gamma_{\beta}
\end{equation}

Or explicitly, $\psi_{\beta}^{0} = \partial_t \Gamma_{\beta}, \; \bm{\psi}_{\beta} = -\bm{\nabla} \Gamma_{\beta}$, which ensures 

\begin{equation} \begin{split} 
\label{unilateral decoupling}
    \bm{\nabla} \wedge \bm{\psi}_{\beta} = -\bm{\nabla} \wedge \bm{\nabla} \Gamma_{\beta} &= 0 \\
    \partial_t \bm{\psi}_{\beta} + \bm{\nabla} \psi_{\beta}^0 = -\partial_t \left( \bm{\nabla} \Gamma_{\beta} \right) + \bm{\nabla} \left( \partial_t \Gamma_{\beta} \right) &= 0
\end{split}
\end{equation}

Then, the system of \crefrange{coupled system scalar alpha}{coupled system vector beta} takes the form:

\begin{equation}
\label{sistema desacoplado psi_4 de psi_1}
    \partial_t \left(
    \begin{array}{c}
    \psi^0_{\alpha} \\
    \partial_t \Gamma_{\beta} \\
    \bm{\psi}_{\alpha} \\
    0
    \end{array} \right)
    + \bm{\nabla} \cdot \left(
    \begin{array}{c}
    \bm{\psi}_{\alpha} \\
    -\bm{\nabla} \Gamma_{\beta} \\
    \bm{\mathsf{T}}_{\alpha} \\
    - I \, \bm{\psi}_{\alpha}
    \end{array} \right)
    = \left(
    \begin{array}{c}
    \phi_{\alpha} \\
    - \phi_{\beta} \\
    - \bm{\phi}_{\alpha} \\
    \bm{\phi}_{\beta}
    \end{array} \right)
\end{equation}

At the cost of restricting the structure of $\psi_{\beta}^\mu$ we get a wave equation for $\Gamma_{\beta}$, however, we free $\psi_{\alpha}^\mu$ in three equations independent of any other component of $\Psi$.

\subsection{Mixed Coupling} \label{Subsection mixed coupling}

As a last notable case, consider instead a multivector for which its components are coupled in the form:

\begin{equation}
\label{Maxwell-like potential}
    \Psi = \left( \partial_t f - \bm{\nabla} f - I \, \bm{\nabla} \wedge \bm{F} \right) \gamma_0 
    + \left( - \bm{\nabla} \cdot \bm{F} + \partial_t \bm{F} \right) I \gamma_0
\end{equation}

where $f = f(x^{\mu})$ is a scalar function and $\bm{F} = \bm{F}(x^{\mu})$ is a vector function. With this structure, the system \crefrange{coupled system scalar alpha}{coupled system vector beta} directly yields the two wave equations

\begin{equation}
\begin{split}
    \label{acoplamiento mixto especial}
    \square f &= \phi_{\alpha} \\
    \square \bm{F} &= \bm{\phi}_{\beta}
\end{split}
\end{equation}

where we have used the identities

\begin{align}
\label{curl of a curl identity}
    \bm{\nabla} \cdot ( I \, \bm{\nabla} \wedge \bm{F} ) &= I \, \bm{\nabla} \wedge (\bm{\nabla} \wedge \bm{F} ) = 0 \\
\label{triple product identity}
    \bm{\nabla} \cdot ( \bm{\nabla} \wedge \bm{F} ) &= \bm{\nabla}^2 \bm{F} - \bm{\nabla} ( \bm{\nabla} \cdot \bm{F} )
\end{align}

Upon comparing with the general case, we see that \cref{Maxwell-like potential,acoplamiento mixto especial} can be put in terms of a potential multivector defined by $f$ and $\bm{F}$:

\begin{equation}
    \begin{split}
        \Psi &= \nabla (f + \bm{F} \, I) \\
        \Phi &= \nabla \Psi = \square ( f + \bm{F} \, I)
    \end{split}
\end{equation}

\section{Symmetries of the Continuity Equation} \label{Sec Symmetries}

From the system of \crefrange{coupled system scalar alpha}{coupled system vector beta} consider the proposal:

\begin{equation}
    \partial_t \psi_{\alpha}^0 + \bm{\nabla} \cdot \bm{\psi}_{\alpha} = \partial_t (\psi_{\alpha}^0)' + \bm{\nabla} \cdot (\bm{\psi}_{\alpha})'
\end{equation}

In other words, how can the functions $\psi_{\alpha}^\mu$ transform in such a way that the structure of its scalar continuity equation remains invariant? Making use of the identity of \cref{curl of a curl identity}, it is found:

\begin{align}
\label{scalar symmetry}
    \psi_{\alpha}^0 &= (\psi_{\alpha}^0)' + \bm{\nabla} \cdot \bm{\Lambda} \\
\label{vector symmetry}
    \bm{\psi}_{\alpha} &= (\bm{\psi}_{\alpha})' - \partial_t \bm{\Lambda} - I \, \bm{\nabla} \wedge \bm{\Theta}
\end{align}

If these symmetry transformations for the scalar continuity are employed in \cref{coupled system vector alpha}, we obtain

\begin{equation}
    \partial_t (\bm{\psi}_{\alpha})' + \bm{\nabla} (\psi_{\alpha}^0)' + I \, \bm{\nabla} \wedge \bm{\psi}_{\beta} - \left[ \square \bm{\Lambda} + \bm{\nabla} \cdot ( I \, \partial_t \bm{\Theta} + \bm{\nabla} \wedge \bm{\Lambda} ) \right] = - \bm{\phi}_{\alpha}
\end{equation}

where we have used identity \cref{triple product identity}. Independently of decoupling or not, let us impose an invariance in the vector continuity \cref{coupled system vector alpha} upon the transformations \crefrange{scalar symmetry}{vector symmetry}:

\begin{equation}
    \label{Field symmetry}
    \partial_t \bm{\psi}_{\alpha} + \bm{\nabla} \psi^0_{\alpha} + I \, \bm{\nabla} \wedge \bm{\psi}_{\beta} = \partial_t (\bm{\psi}_{\alpha})' + \bm{\nabla} (\psi^0_{\alpha})' + I \, \bm{\nabla} \wedge \bm{\psi}_{\beta}
\end{equation}

Which then delivers the condition

\begin{equation}
    \label{wave of Lambda}
    \square \bm{\Lambda} = - I\, \bm{\nabla} \wedge ( \partial_t \bm{\Theta} - I \, \bm{\nabla} \wedge \bm{\Lambda} )
\end{equation}

This allows us to define

\begin{equation}
    \label{Faraday-Maxwell}
    \partial_t \bm{\Theta} - I \, \bm{\nabla} \wedge \bm{\Lambda} = \bm{\Omega}
\end{equation}

By applying the $\partial_t$ operator to \cref{Faraday-Maxwell}, the $I \, \bm{\nabla} \wedge$ operator to \cref{vector symmetry}, and combining the two expressions we obtain

\begin{equation}
    \label{wave of Theta}
    \square \bm{\Theta} = \partial_t \bm{\Omega} - \bm{\nabla} ( \bm{\nabla} \cdot \bm{\Theta} ) - I \, \bm{\nabla} \wedge \left[ \bm{\psi}_\alpha - (\bm{\psi}_\alpha)' \right]
\end{equation}

Finally, it is convenient to define

\begin{equation}
    \label{Gauss magnetism}
    \bm{\nabla} \cdot \bm{\Theta}  = -\omega
\end{equation}

Thus, grouping all the results into matrix-notation continuity systems:

\begin{equation}
\label{Maxwell Equations}
    \partial_t \left(
    \begin{array}{c}
    0 \\
    0 \\
    \bm{\Lambda} \\
    \bm{\Theta}
    \end{array} \right)
    + \bm{\nabla} \cdot \left(
    \begin{array}{c}
    \bm{\Lambda} \\
    \bm{\Theta} \\
    I \, \bm{\Theta} \\
    - I \, \bm{\Lambda}
    \end{array} \right)
    = \left(
    \begin{array}{c}
    \psi^0_{\alpha} - (\psi^0_{\alpha})' \\
    - \omega \\
    - \big[ \bm{\psi}_{\alpha} - (\bm{\psi}_{\alpha})' \big] \\
    \bm{\Omega}
    \end{array} \right)
\end{equation}

\begin{equation}
\label{Maxwell conservation}
    \partial_t \left(
    \begin{array}{c}
    \psi^0_{\alpha} - (\psi^0_{\alpha})' \\
    \omega \\
    \end{array} \right)
    + \bm{\nabla} \cdot \left(
    \begin{array}{c}
    \bm{\psi}_{\alpha} - (\bm{\psi}_{\alpha})' \\
    \bm{\Omega} \\
    \end{array} \right)
    = \square \left(
    \begin{array}{c}
    0 \\
    0 \\
    \end{array} \right)
\end{equation}

\begin{equation}
\label{Maxwell Wave Equations}
    \partial_t \left(
    \begin{array}{c}
    \bm{\psi}_{\alpha} - (\bm{\psi}_{\alpha})' \\
    \bm{\Omega} \\
    \end{array} \right)
    + \bm{\nabla} \left(
    \begin{array}{c}
    \psi^0_{\alpha} - (\psi^0_{\alpha})' \\
    \omega \\
    \end{array} \right)
    - I \, \bm{\nabla} \wedge \left(
    \begin{array}{c}
    - \bm{\Omega} \\
    \bm{\psi}_{\alpha} - (\bm{\psi}_{\alpha})'
    \end{array} \right)
    = \square \left(
    \begin{array}{c}
    - \bm{\Lambda} \\
    \bm{\Theta}
    \end{array} \right)
\end{equation}

Ergo, by comparing with the general structures \cref{tensor continuity psi,tensor continuity phi}, it follows that if

\begin{align}
    \label{Faraday Bivector}
    \bm{Q} &= \bm{\Lambda} + \bm{\Theta} \, I \\
    \label{current-like multivector}
    M &= \left[ \psi^0_{\alpha} - (\psi^0_{\alpha})' + \bm{\psi}_{\alpha} - (\bm{\psi}_{\alpha})' \right]\gamma_0 + (\omega + \bm{\Omega} ) I \gamma_0
\end{align}

then \cref{Maxwell Equations,Maxwell conservation,Maxwell Wave Equations} are nothing but

\begin{align}
    \label{Maxwell-like equation}
    \nabla \bm{Q} &= M \\
    \label{Waves of the Faraday}
    \square \bm{Q} &= \nabla M
\end{align}

\subsection{Free Waves} \label{Subsection free waves}

The condition for homogeneous wave equations for the fields $\bm{\Lambda}$ and $\bm{\Theta}$,

\begin{equation}
    \square \bm{\Lambda} = \square \bm{\Theta} = 0
\end{equation}

on one hand imposes the invariance of \cref{coupled system vector beta},

\begin{equation}
    - I \, \bm{\nabla} \wedge \bm{\psi}_\alpha = - I \, \bm{\nabla} \wedge (\bm{\psi}_\alpha)'
\end{equation}

and on the other hand the restrictions on $\langle M \rangle_3$:

\begin{equation}
\begin{split}
    \bm{\nabla} \wedge \bm{\Omega} &= 0 \\
    \partial_t \bm{\Omega} + \bm{\nabla} \omega &= 0
\end{split}
\end{equation}

But from \cref{Subsection unilateral decoupling} these are just the conditions for the unilateral decoupling of $\langle M \rangle_3$ with respect to $\langle M \rangle_1$. Thus, the pseudovector of $M$ takes the form

\begin{equation}
    \langle M \rangle_3 = ( \partial_t \zeta - \bm{\nabla}\zeta ) I \gamma_0
\end{equation}

\subsection{Potentials}

\Cref{Subsection mixed coupling,Maxwell-like equation,Waves of the Faraday} allow us to consider the equations:

\begin{align}
\label{Potential equation}
    \nabla P &= \bm{Q} \\
\label{wave for the potentials}
    \square P &= \nabla \bm{Q} = M
\end{align}

From structure \cref{tensor continuity psi} we see that the bivector $\bm{Q}$ is completely characterised by a potential multivector of the form

\begin{equation}
    \label{Potential}
    P = (p_1^0 + \bm{p}_1)\gamma_0 + (p_4^0 + \bm{p}_4)I\gamma_0
\end{equation}

Therefore, \cref{Potential equation,wave for the potentials} in explicit form are

\begin{equation}
\label{potentials conservation}
    \partial_t \left(
    \begin{array}{c}
    p_1^0 \\
    p_4^0 \\
    \end{array} \right)
    + \bm{\nabla} \cdot \left(
    \begin{array}{c}
    \bm{p}_1 \\
    \bm{p}_4 \\
    \end{array} \right)
    = \left(
    \begin{array}{c}
    0 \\
    0 \\
    \end{array} \right)
\end{equation}

\begin{equation}
    \partial_t \left(
    \begin{array}{c}
    \bm{p}_1 \\
    \bm{p}_4 \\
    \end{array} \right)
    + \bm{\nabla} \left(
    \begin{array}{c}
    p_1^0 \\
    p_4^0 \\
    \end{array} \right)
    -I \bm{\nabla} \wedge \left(
    \begin{array}{c}
    - \bm{p}_4 \\
    \bm{p}_1
    \end{array} \right)
    = \left(
    \begin{array}{c}
    - \bm{\Lambda} \\
    \bm{\Theta}
    \end{array} \right)
\end{equation}

\begin{equation}
    \square \left(
    \begin{array}{c}
    p_1^0 \\
    - p_4^0 \\
    - \bm{p}_1 \\
    \bm{p}_4
    \end{array} \right)
    = \partial_t \left(
    \begin{array}{c}
    0 \\
    0 \\
    \bm{\Lambda} \\
    \bm{\Theta}
    \end{array} \right)
    + \bm{\nabla} \cdot \left(
    \begin{array}{c}
    \bm{\Lambda} \\
    \bm{\Theta} \\
    I \, \bm{\Theta} \\
    - I \, \bm{\Lambda}
    \end{array} \right)
    = \left(
    \begin{array}{c}
    \psi_1^0 - (\psi_1^0)' \\
    - \omega \\
    - \big[ \bm{\psi}_1 - (\bm{\psi}_1)' \big] \\
    \bm{\Omega}
    \end{array} \right)
\end{equation}

Imposing the decoupling of $\langle P \rangle_3$ with respect to $\langle P \rangle_1$ through \cref{psi_4 desacoplamiento} permits to express the fields in terms of the vector potential exclusively,

\begin{equation}
    \begin{split}
        \bm{\Lambda} &=  - \bm{\nabla} p_1^0 - \partial_t \bm{p}_1 \\
        \bm{\Theta} &= - I \, \bm{\nabla} \wedge \bm{p}_1
    \end{split}    
\end{equation}

\subsection{Poynting}

By the structure of \cref{Faraday Bivector}, the Poynting multivector associated with the field $\bm{Q}$ has the structure described in \cref{Poynting Cases}. That is,

\begin{equation}
\label{Poynting EM}
    S_{\bm{Q}} = \frac{1}{2} \Big( \bm{\Lambda} \cdot \bm{\Lambda} + \bm{\Theta} \cdot \bm{\Theta} \Big) \gamma_0 - I \, \bm{\Lambda} \wedge \bm{\Theta} \, \gamma_0
\end{equation}

Further, by the structure of \cref{Potential} we may also construct a Poynting multivector for the potential $P$. Its general form corresponds to that described in \cref{S for vector Psi}. We show the special case where $p_4^\mu$ is null:

\begin{equation}
    S_{P} = \frac{1}{2} \left[ (p^0_{1})^2 + \bm{p}_1 \cdot \bm{p}_1 \right] \gamma_0 + p^0_{1} \bm{p}_1 \gamma_0
\end{equation}

The respective continuity equations are:

\begin{align}
    \partial_t \Big[ \frac{1}{2} \Big( \bm{\Lambda}^2 + \bm{\Theta}^2 \Big) \Big] + \bm{\nabla} \cdot \Big( - I \, \bm{\Lambda} \wedge \bm{\Theta} \Big) &= \langle W \rangle_0 \label{energy EM conservation} \\
    \partial_t \Big[ \frac{1}{2} \left( p^0_{1}p^0_{1} + \bm{p}_1 \cdot \bm{p}_1 \right) \Big] + \bm{\nabla} \cdot \Big( p^0_{1} \bm{p}_1 \Big) &= \langle U \rangle_0
\end{align}

\subsection{Discussion}

We have found that the invariance of the system of continuity equations for a multivector $\langle \Psi \rangle_1$ due to symmetry transformations naturally allows for the construction of a system of eqs. with the structure of Maxwell's field equations\footnote{To express in Gibbs-Heaviside's vector calculus notation: $ \bm{\nabla} \cdot ( I \bm{F} ) = I \, \bm{\nabla} \wedge \bm{F} = - \bm{\nabla} \times \bm{F}$.}, \cref{Maxwell Equations}, for the bivector field $\bm{Q}$ defined by \cref{Faraday Bivector}. Because of the symmetric nature in the system of \crefrange{coupled system scalar alpha}{coupled system vector beta} for $\langle \Psi \rangle_1 + \langle \Psi \rangle_3$ and $\langle \Psi \rangle_0 + \langle \Psi \rangle_2 + \langle \Psi \rangle_4$, all of these relations hold if $M$ has a bivector character and $Q$ a vector character. The emergence of the new pseudovector term $\langle M \rangle_3 = \left( \omega + \bm{\Omega} \right) I \gamma_0$ exists in independence of the pseudovector part $\langle \Psi \rangle_3 = \left( \psi^0_\beta + \bm{\psi}_\beta \right) I \gamma_0$ of $\Psi$,

\begin{equation}
    M = \langle \Psi \rangle_1 - \langle \Psi' \rangle_1 + \langle M \rangle_3
\end{equation}

The symmetries do admit $\langle \Psi \rangle_3 = \langle M \rangle_3$, although this does not seem to be a requirement but a special case. The whole system is obtained by only the symmetries of \cref{coupled system scalar alpha,coupled system vector alpha} insofar as the invariance of \cref{coupled system vector beta} emerges as a condition for the existence of free waves for the fields. The vector part $\langle M \rangle_1 = \langle \Psi \rangle_1 - \langle \Psi' \rangle_1$ may be regarded as a displacement in the transported physical quantity with a fixed coordinate system. 

If, by hypothesis, $\langle \Psi \rangle_1 \to J \gamma_0 = \rho + \bm{j}$ in \cref{current-like multivector} is the space-time electric current density, then it can be immediately identified $\bm{Q} \to \bm{F} = \bm{E} + I \bm{B}$ from \cref{Faraday Bivector} as Faraday's electromagnetic field, $\langle P \rangle_1 \to A \gamma_0 = \phi + \bm{A}$ from \cref{Potential} as the space-time electromagnetic potential, and $S_1^0 \to u = \frac{1}{2} \Big( \bm{E}^2 + \bm{B}^2 \Big)$, $\bm{S}_1 \to \bm{s} = \bm{E} \times \bm{B}$ of \cref{Poynting EM} as the electromagnetic energy density and Poynting vector, respectively. $\langle M \rangle_3$ may be identified as the space-time magnetic current density, and $\langle P \rangle_3$ as a new space-time magnetic potential. Correspondingly, \cref{Maxwell conservation} represents the electric and magnetic charge conservation, \cref{potentials conservation} the Lorenz gauge, and \cref{energy EM conservation} the conservation of electromagnetic energy\footnote{In STA's treatment of electromagnetic theory it is well-known that the expansion of \cref{Dirac's current continuity} and $\frac{1}{2} \Big( \widetilde{\Phi} \Psi + \widetilde{\Psi} \Phi \Big)$ is connected to Poynting's theorem and Lorentz force \cite{Vold1993}.}. The fact that the condition of free waves for the electric and magnetic fields implies the decoupling of the magnetic current density $\langle M \rangle_3$ with respect to the electric current density $J$ (\cref{Subsection free waves}) and also that the magnetic charge density is negative, \cref{Gauss magnetism}, may shed light to the nature of magnetic monopoles and their experimental elusiveness so far.

If $\langle \Psi \rangle_1 \to J \gamma_0 = n + \bm{j}$ is the turbulent or fluid-mechanical `charge' density, then $\bm{\Lambda} \to \bm{L}$ is Lamb's vector, $\bm{\Omega} \to \bm{w}$ is the vorticity, $\bm{p}_1 \to \bm{v}$ is the fluid's velocity, $p_1^0 \to h$ is the enthalpy per unit mass or equivalently $p_1^0 \to \varphi$ Bernoulli's energy function, and \cref{Maxwell Equations} are Maxwell's field equations for a compressible fluid \cite{Demir2017, Sen2023}. But, of course, the results found that Maxwell's equations are a consequence of the continuity equation's symmetries are not limited to either electrodynamics or hydrodynamics.

\section{Diffusion Equation} \label{Sec Diffusion}

To conclude, consider the case where one of the relative vectors satisfies Fick's laws of diffusion. E.g.,

\begin{equation}
    \label{Diffusion}
    \bm{\psi}_{\alpha} = - D \, \bm{\nabla} \psi^0_{\alpha}
\end{equation}

where $D = D\left( \psi^0_{\alpha}; x^\mu \right)$ is called the diffusion coefficient. With \cref{Diffusion}, the continuity system of \crefrange{coupled system scalar alpha}{coupled system vector beta} takes the form:

\begin{align}
    \label{convection diffusion system 1}
    \partial_t \psi_{\alpha}^0 - \bm{\nabla} \cdot ( D \, \bm{\nabla} \psi^0_{\alpha} ) &= \phi_{\alpha} \\
    \partial_t \psi_{\beta}^0 + \bm{\nabla} \cdot \bm{\psi}_{\beta} &= - \phi_{\beta} \\
    \left( 1 - \partial_t D \right) \bm{\nabla} \psi^0_{\alpha} - D \, \bm{\nabla} \left( \partial_t \psi^0_{\alpha} \right) + I \, \bm{\nabla} \wedge \bm{\psi}_{\beta} &= - \bm{\phi}_{\alpha} \\
    \label{convection diffusion system 4}
    \partial_t \bm{\psi}_{\beta} + \bm{\nabla} \psi^0_{\beta} + I \, \bm{\nabla} D \wedge \bm{\nabla} \psi^0_{\alpha} &= \bm{\phi}_{\beta}
\end{align}

\Cref{Diffusion} establishes a diffusion equation for $\psi^0_{\alpha}$ in \cref{convection diffusion system 1}. However, the solution and structure of this equation are conditioned by the rest of the equations in the system.

\subsection{Constant Diffusion Coefficient} \label{Subsec const coeff}

Consider the case $D = \text{constant}$. For this condition, the system of \crefrange{convection diffusion system 1}{convection diffusion system 4} reduces to:

\begin{equation}
\begin{split}
    \label{heat system}
    \partial_t \psi_{\alpha}^0 - D \, \bm{\nabla}^2 \psi^0_{\alpha} &= \phi_{\alpha} \\
    \partial_t \psi_{\beta}^0 + \bm{\nabla} \cdot \bm{\psi}_{\beta} &= - \phi_{\beta} \\
    \bm{\nabla} \left( \psi^0_{\alpha} - D \, \partial_t \psi^0_{\alpha} \right) + I \, \bm{\nabla} \wedge \bm{\psi}_{\beta} &= - \bm{\phi}_{\alpha} \\
    \partial_t \bm{\psi}_{\beta} + \bm{\nabla} \psi^0_{\beta} &= \bm{\phi}_{\beta}
\end{split}
\end{equation}

As expected, a constant diffusion coefficient yields a heat equation for $\psi_{\alpha}^0$. The remaining equations are still coupled with it, so let us inspect the case for which $\bm{\psi}_{\beta}$ is subjected to the imposition of \cref{desacoplamiento simultaneo} such that \cref{heat system} takes the form:

\begin{equation}
\begin{split}
    \partial_t \psi_{\alpha}^0 - D \, \bm{\nabla}^2 \psi^0_{\alpha} &= \phi_{\alpha} \\
    \partial_t \psi_{\beta}^0 + \bm{\nabla}^2 \Gamma_{\beta} &= - \phi_{\beta} \\
    \bm{\nabla} \left( \psi^0_{\alpha} - D \, \partial_t \psi^0_{\alpha} \right) &= - \bm{\phi}_{\alpha} \\
    \bm{\nabla} \left( \psi^0_{\beta} + \partial_t \Gamma_{\beta} \right) &= \bm{\phi}_{\beta}
\end{split}
\end{equation}

Decoupling of $\psi^\mu_{\alpha}$ and $\psi^\mu_{\beta}$ is indeed achieved. However, it can be seen that the absence of external sources, $\nabla \Psi = 0$, represents a separation of space and time variables for $\psi^0_{\alpha}$. That is, $\Phi = 0$ implies

\begin{equation}
\begin{split}
    \label{alfa decoupling diffusion}
    \psi^0_{\alpha} - D \, \partial_t \psi^0_{\alpha} &= \text{constant} \\
    \psi^0_{\beta} + \partial_t \Gamma_{\beta} &= \text{constant}
\end{split}
\end{equation}

and thus

\begin{align}
    \label{Helmholtz}
    \left( \bm{\nabla}^2 - D^{-2} \right) \psi^0_{\alpha} &= \text{const.} \\
    \label{wave of gamma beta}
    \square \, \Gamma_{\beta} &= 0
\end{align}

For a pseudoscalar diffusion coefficient, $D = (D')^{-1} I$, \cref{Helmholtz} takes the structure of a Helmholtz equation: $\left( \bm{\nabla}^2 + D'^{2} \right) \psi^0_{\alpha} = \text{const}$. The separation of variables for $\psi^0_{\alpha}$ constitutes by itself a transformation that decouples the system, as can be easily verified by imposing \textit{a priori} \cref{alfa decoupling diffusion} to the system of \cref{heat system}. Furthermore, the decoupling-by-gradient hypothesis for $\bm{\psi}_{\beta}$ may also adopt a diffusion structure by considering a multivector $\Psi = (\psi_\alpha - D_\alpha \bm{\nabla}\psi_\alpha) + \left( \psi_\beta - D_\beta \bm{\nabla}\psi_\beta \right) I$, where $D_{\alpha, \beta}$ are both constants. \Cref{nabla psi = phi} yields the `secondly decoupled' system:

\begin{equation}
\begin{split}
    \partial_t \psi_{\alpha,\beta}  - D_{\alpha, \beta} \, \bm{\nabla}^2 \psi_{\alpha, \beta} &= \pm \phi_{\alpha, \beta} \\
    \bm{\nabla} \left( \psi_{\alpha,\beta} - D_{\alpha, \beta} \, \partial_t \psi_{\alpha,\beta} \right) &= \mp \bm{\phi}_{\alpha,\beta}
\end{split}
\end{equation}

Once again, the absence of external sources $\Phi = 0$---or specifically the absence of vector field sources $\bm{\phi}_{\alpha, \beta} = 0$---reduces the system into two Helmholtz-like equations, \cref{Helmholtz}.

\subsection{Non-Constant Diffusion Coefficient}

Let us decouple $\psi^\mu_{\beta}$ from the general diffusion system of \crefrange{convection diffusion system 1}{convection diffusion system 4}:

\begin{align}
    \label{nonconstant diffusion 1}
    \partial_t \psi_{\alpha}^0 - \bm{\nabla} \cdot ( D \, \bm{\nabla} \psi^0_{\alpha} ) &= \phi_{\alpha} \\
    \label{nonconstant diffusion 2}
    \left( 1 - \partial_t D \right) \bm{\nabla} \psi^0_{\alpha} - D \, \bm{\nabla} \left( \partial_t \psi^0_{\alpha} \right) &= - \bm{\phi}_{\alpha} \\
    \label{nonconstant diffusion 3}
    I \, \bm{\nabla} D \wedge \bm{\nabla} \psi^0_{\alpha} &= \bm{\phi}_{\beta}
\end{align}

Consider the case where the diffusion coefficient is an explicit function of $\psi^0_{\alpha}$ but not on the space-time coordinates, $D = D\left( \psi^0_{\alpha} \right)$:

\begin{equation}
\begin{split}
\label{irrotacional}
    \partial_t D &= \frac{\partial D}{\partial \psi^0_{\alpha}} \partial_t \psi^0_{\alpha} \\
    \bm{\nabla} D &= \frac{\partial D}{\partial \psi^0_{\alpha}} \bm{\nabla} \psi^0_{\alpha}
\end{split}
\end{equation}

Then, from \crefrange{nonconstant diffusion 1}{nonconstant diffusion 3} the following non-linear system is obtained

\begin{equation}
\begin{split}
\label{Difussion explicit}
    \partial_t \psi_{\alpha}^0 - \bm{\nabla} \cdot ( D \, \bm{\nabla} \psi^0_{\alpha} ) &= \phi_{\alpha} \\
    \bm{\nabla} \psi^0_{\alpha} - \frac{\partial D}{\partial \psi^0_{\alpha}} \partial_t \psi^0_{\alpha} \bm{\nabla} \psi^0_{\alpha} - D \, \bm{\nabla} \left( \partial_t \psi^0_{\alpha} \right) &= - \bm{\phi}_{\alpha}
\end{split}
\end{equation}

Structure $D = D \left( \psi^0_{\alpha} \right)$ by virtue of \cref{irrotacional} implies that the diffusion flux is a conservative vector field, which further enables the introduction of a potential gradient:

\begin{equation}
     \bm{\nabla}{\vartheta} = D \, \bm{\nabla} \psi^0_{\alpha}
\end{equation}

Ergo, system \cref{Difussion explicit} can also be expressed as

\begin{equation}
\begin{split}
    \partial_t \psi_{\alpha}^0 - \bm{\nabla}^2 \vartheta &= \phi_{\alpha} \\
    \bm{\nabla} \left( \psi^0_{\alpha} - \partial_t \vartheta \right) &= - \bm{\phi}_{\alpha}
\end{split}
\end{equation}

Once again, in the absence of external sources $\psi^0_{\alpha} - \partial_t \vartheta = \text{const}$, and therefore

\begin{equation}
    \square \, \vartheta = 0
\end{equation}

\subsubsection{Interpretation}

Consider the system of \crefrange{nonconstant diffusion 1}{nonconstant diffusion 3} product of $\nabla \left( \psi^0_\alpha - D \bm{\nabla} \psi^0_\alpha \right) = \left( \phi^0_\alpha + \bm{\phi}_\alpha + I \bm{\phi}_\beta \right) \gamma_0$. \Cref{nonconstant diffusion 2,nonconstant diffusion 3} condition the function $\psi^0_\alpha$ in such a way that for constant or explicitly-dependent coefficients, the absence of external sources implies that the diffusion structure of \cref{nonconstant diffusion 1} is lost. This situation can be understood from a perspective of bi-dependence between the components of $\Psi$ and $\Phi$. From \cref{nonconstant diffusion 1}, the density/concentration $\psi^0_\alpha$ can be determined in terms of the scalar source $\phi^0_\alpha$ and the diffusive coefficient $D$. Then, $(\bm{\phi}_\alpha + I \bm{\phi}_\beta)$ may be conceived as an external vector-field source \textit{defined} by \cref{nonconstant diffusion 2,nonconstant diffusion 3} in terms of $\psi^0_\alpha$ and $D$. From the constant and explicitly-dependent coefficient cases, we saw that if this field is null then the diffusive structure is transformed into Helmholtz-like and wave structures. Therefore, the external field $(\bm{\phi}_\alpha + I \bm{\phi}_\beta)$ admits the interpretation of being the source that produces itself the diffusion phenomena.

\section{Conclusion}

In synthesis, we have shown that for a generalised STA multivector, equation $\nabla \Psi = \Phi$ yields a system of eight continuity equations---four of scalar character and four of vector character. From this, we constructed the system $\square \Psi = \nabla \Phi$ where each wave equation for the components of $\Psi$ plays the role of the source for the continuity system of $\Phi$. With the same method for which the continuity equation is derived from Schrödinger and Dirac equations, we have defined the Poynting multivector associated with the density of the generalised multivector $\Psi$, together with its continuity system. These systems allow the identification of the coupled structures $\langle \Psi \rangle_1 + \langle \Psi \rangle_3$ and $\langle \Psi \rangle_0 + \langle \Psi \rangle_2 + \langle \Psi \rangle_4$. The cost of complete decoupling for either structure is a transition to a system of higher-order equations, such as the wave system. Faced with this, we have determined three cases of decoupling by imposing restrictions on the structure of at least one of the functions involved in the continuity equations: if both functions are APS gradients; if one of them is a space-time gradient; and, considering instead a mixed coupling, a potential multivector structure was constructed for which the continuity system yields directly a wave system. Symmetry transformations that make the (scalar) continuity equation invariant are found in terms of vector fields. By imposing that the continuity equations system be invariant under these transformations, a system of field and wave equations with the structure of Maxwell's equations was derived. This Maxwellian system along with an application of the continuity, wave, and Poynting multivector systems permitted the construction of fields and potentials directly connected to the generalised STA multivector: $\square P = \nabla \bm{Q} = \langle \Psi \rangle_1 - \langle \Psi' \rangle_1$. The results found are consistent with the classical electromagnetic theory and with hydrodynamics from fluid mechanics. When one of the functions in the continuity system satisfies Fick's laws, a diffusion equation arises coupled with the remaining continuity equations of the system. It was found that for a decoupled system with a constant or explicitly-dependent diffusion coefficient, the absence of external sources implies a loss in the diffusion equation structure being transformed to a Helmholtz-like structure and a wave system. For both cases considered, the external vector-field source can be defined by the complementary equations in the system in terms of the density and diffusion coefficient, and admit the interpretation of the diffusion phenomena producer.

\printbibliography

\end{document}